\documentstyle[prd,aps]{revtex}
\begin{document}
\draft
\input epsf
\twocolumn[\hsize\textwidth\columnwidth\hsize\csname
@twocolumnfalse\endcsname

\title{GUT Baryogenesis after Preheating}
\author{Edward W. Kolb,$^{(1,2)}$ Andrei Linde,$^{(3)}$ and Antonio
Riotto$^{(1)}$}
\address{$^{(1)}${\it NASA/Fermilab Astrophysics Center, \\ Fermilab
National Accelerator Laboratory, Batavia, Illinois~~60510-0500}}
\address{$^{(2)}${\it Department of Astronomy and Astrophysics,
Enrico Fermi Institute,\\
University of Chicago, Chicago, Illinois~~60637-1433}}
\address{$^{(3)}${\it Department of Physics, Stanford University,
Stanford, California~~94305-4060}}
\date{June 3, 1996}
\maketitle
\begin{abstract}
At the end of inflation the universe is frozen in a near zero-entropy
state with energy density in a coherent scalar field and must be
``defrosted'' to produce the observed entropy and baryon number.  We
propose that the baryon asymmetry is generated by the decay of
supermassive Grand Unified Theory (GUT) bosons produced non-thermally
in a preheating phase after inflation.  We show that
baryogenesis is possible for an inflaton masses of order $10^{13}$GeV
and a GUT Higgs boson mass of order $10^{14}$GeV, thus solving many
drawbacks facing GUT baryogenesis in the old reheating scenario.
\end{abstract}
\pacs{PACS: 98.80.Cq    \hskip 1 cm FERMILAB--Pub--96/133-A,~
SU-ITP-96-24 \hskip 1 cm
 hep-ph/9606260}
\vskip2pc]

In models of slow-roll inflation \cite{las,chaotic}, the universe is
dominated by the potential energy density of a scalar field known as
the {\it inflaton}.  Inflation ends when the kinetic energy density of
the inflaton becomes larger than its potential energy density.  At
this point the universe might be said to be frozen: any initial
entropy in the universe was inflated away, and the only energy was in
cold, coherent motions of the inflaton field.  Somehow this frozen
state must be transformed to a high-entropy hot universe by
transferring energy from the inflaton field to radiation.  This
process is usually called reheating, which may well be a misnomer
since there is no guarantee that the universe was hot before
inflation.  Since we are confident that the universe was
frozen at the end of inflation, perhaps ``defrosting'' is a better
description of the process of converting inflaton coherent energy into
entropy.

In the old reheating (defrosting) scenario \cite{old}, the inflaton
field $\phi$ is assumed to oscillate coherently about the minimum of
the inflaton potential until the age of the universe is equal to the
lifetime of the inflaton. Then the inflaton decays, and the decay
products thermalize to a temperature $T_F\simeq 10^{-1}
\sqrt{\Gamma_\phi M_{\rm P}}$, where $\Gamma_\phi$ is the inflaton
decay width, and $M_{\rm P}\sim 10^{19}$ GeV is the Planck mass.  In
the simple chaotic inflation model we study the potential is assumed
to be $V(\phi) = M_\phi^2\phi^2/2$, with $M_\phi\sim 10^{13}$GeV in
order to reproduce the observed temperature anisotropies in the
microwave background.  If we write $\Gamma_\phi=\alpha_\phi M_\phi$,
then $T_F\simeq 10^{15}\sqrt{\alpha_\phi}$ GeV \cite{ktbook}.

In supergravity-inspired scenarios, gravitinos have a mass of order a
TeV and a decay lifetime on the order of $10^5$s. If gravitinos are
overproduced after inflation and decay after the epoch of
nucleosynthesis, they would modify the successful predictions of
big-bang nucleosynthesis.  This can be avoided if the temperature
$T_F$ is smaller than about $10^{11}$ GeV (or even less, depending on
the gravitino mass) \cite{limit}, which implies $\alpha_\phi {\
\lower-1.2pt\vbox{\hbox{\rlap{$<$}\lower5pt\vbox{\hbox{$\sim$}}}}\ }
10^{-8}$.

In addition to entropy, the baryon asymmetry must be created after
inflation.  There are serious obstacles facing any attempt to generate
a baryon asymmetry in an inflationary universe through the decay of
baryon number ($B$) violating bosons of Grand Unified Theories
\cite{review}.  The first problem is that $B$ violation through
sphaleron transitions are expected to be fast at high temperatures,
and would erase any preexisting baryon asymmetry produced at the GUT
scale \cite{sh} unless there is a non vanishing value of $B-L$.  But a
natural way to overcome this problem is to adopt a GUT like $SO(10)$,
where an asymmetry in $(B-L)$ may be generated.

A more serious problem is the low value of $T_F$ in the old scenario.
Since the unification scale is expected to be of order $10^{16}$GeV,
$B$ violating gauge and Higgs bosons (referred to generically as
``$X$'' bosons) probably have masses greater than $M_\phi$, and it
would be kinematically impossible to produce them directly in $\phi$
decay.\footnote{Gauge bosons have masses comparable to the unification
scale, while $B$ violating Higgs bosons may have a mass a few orders
of magnitude less. For example, in $SU(5)$ there are $B$ violating
``Higgs'' bosons in the five-dimensional representation that may have
a mass as small as $10^{14}$GeV. In fact, these Higgs bosons are a
more likely than gauge bosons to produce a baryon asymmetry since it
is easier to arrange the requisite CP violation in the Higgs decay
\cite{KW,FOT,CP}.  Furthermore, if $T_F$ is less than $10^{11}$GeV,
$X$ bosons will be exponentially rare in the thermal background after
inflation.}

However, it has been recently realized \cite{explosive,richie} that
reheating may differ significantly from the above simple picture.  In
the first stage of reheating, which was called ``preheating''
\cite{explosive}, nonlinear quantum effects may lead to an extremely
effective dissipational dynamics and explosive particle production
even when single particle decay is kinematically forbidden. Particles
can be produced in a regime of a broad parametric resonance, and it is
possible that a significant fraction of the energy stored in the form
of coherent inflaton oscillations at the end of inflation is released
after only a dozen oscillation periods.

This Letter demonstrates that preheating may play an extremely
important role for the GUT generation of the baryon asymmetry, as
first suggested in \cite{explosive,KLSSR} (see also \cite{y,tkachev}).
Indeed, we will show that the baryon asymmetry can be produced
efficiently just after the preheating era, thus solving many of the
problems that GUT baryogenesis had to face in the old picture of
reheating.

There are several different ways to resurrect GUT baryogenesis. The
simplest way is to take into account that if particles produced at
preheating can rapidly decay, then the reheating temperature may be
very large, which may lead to the standard thermal production of
superheavy $X$ particles.  However, if the products of parametric
resonance are capable of an instantaneous decay and thermalization,
then the parametric resonance never happens. Out of all possible ways
of development of parametric resonance, Nature chooses only those
which do not lead to an instantaneous thermalization.  In general, it
does not preclude sufficiently high reheating temperature and
subsequent baryogenesis, which may appear if the bosons produced at
preheating decay and thermalize fast, but not fast enough to destroy
the resonance. However, by assuming a thermal mechanism for $X$ boson
production one is loosing the advantage of the non-equilibrium nature
of preheating, which may allow for a direct non-thermal creation of
$X$ bosons.

Indeed, preheating occurs because the interaction terms of the type of
$\lambda_\phi \phi^2 |X|^2$ gives the oscillating contribution
$\lambda\phi^2(t)$ to the mass squared of bosons interacting with the
inflaton field.  This leads to a broad parametric resonance in an
expanding universe for $\lambda_\phi\bar\phi^2 > M^2_\phi$, where
$\bar\phi$ is the amplitude of the oscillating inflaton field
\cite{explosive}. The first stage of reheating does not extract all
the initial energy of the inflaton field.  As the amplitude of the
oscillations of the inflaton field decreases, one leaves the resonance
regime, and particle production ceases \cite{explosive}.

A crucial observation for baryogenesis is that even particles with
mass larger than the inflaton mass may be produced during preheating.
While previous studies of preheating concentrated on creation of light
particles, the results can be easily generalized to supermassive particles as
well.

Following \cite{explosive}, during preheating quantum fluctuations of
the $X$ field with   momentum $\vec{k}$ obey the Mathieu
equation: $ X_k'' + [A(k) - 2q\cos2z]X_k =0$, where $q
= \lambda_\phi \phi^2 / 4 M_\phi^2$, $A(k) = (k^2 + M_X^2) / M_\phi^2 + 2q$,
and primes denotes differentiation with respect to
$z=M_\phi t$.  Particle production in the broad resonance regime occurs
above the line $A = 2 q$.  The width of the instability strip scales as
$q^{1/2}$
for large $q$, independent of the $X$ mass.    The condition for broad
resonance, $A-2q {\
\lower-1.2pt\vbox{\hbox{\rlap{$<$}\lower5pt\vbox{\hbox{$\sim$}}}}\ } q^{1/2}$
\cite{explosive}, becomes $(k^2 + M^2_X)/M_\phi^2 {\
\lower-1.2pt\vbox{\hbox{\rlap{$<$}\lower5pt\vbox{\hbox{$\sim$}}}}\ }
\lambda_\phi^{1/2} \phi / 2M_\phi$, which yields $E_X^2 = {k^2 + M^2_X
} {\ \lower-1.2pt\vbox{\hbox{\rlap{$<$}\lower5pt\vbox{\hbox{$\sim$}}}}\ }
\lambda_\phi^{1/2} \phi M_\phi / 2$.

Therefore the typical energy of $X$ bosons produced in preheating is $E_X^2
\sim
\lambda_\phi^{1/2} \bar\phi  M_\phi$ \cite{explosive}. At the end of the broad parametric resonance this equation somewhat changes because of the backreaction of produced particles. The  resulting estimate for the amplitude of perturbations and for the typical energy of particles at the end of the broad resonance regime for $M_\phi \sim 10^{-6} M_{\rm P}$ is:  $\langle X^2\rangle^{1/2} \sim 10^{-1} \lambda_\phi^{-1/4} \sqrt {M_\phi M_{\rm P}} \sim \lambda_\phi^{-1/4} 10^{15} $ GeV,   $E_X \sim 10^{-1} \lambda_\phi^{1/4}\sqrt { M_\phi M_{\rm P}} \sim \lambda_\phi^{1/4} 10^{15} $ GeV   \cite{explosive,KLSSR}.   $X$ bosons can
be produced by the broad parametric resonance for $E_X > M_X$, i.e.,
for $M_X < \lambda_\phi^{1/4} 10^{15}$ GeV. For $\lambda_\phi \sim 1$
one would have copious production of particles as heavy as $10^{15}$
GeV, i.e., 100 times greater than the inflaton mass. In what follows
we will consider the model with $M_X = 10^{14}$ GeV. Such particles
can be produced by parametric resonance for $\lambda_\phi {\
\lower-1.2pt\vbox{\hbox {\rlap{$>$}\lower5pt\vbox{\hbox {$\sim$}}}}\
} 10^{-3} - 
10^{-4}$ \cite{KT}. The only problem here is that for $\lambda_\phi {\
\lower-1.2pt\vbox{\hbox {\rlap{$>$}\lower5pt\vbox{\hbox {$\sim$}}}}\
}10^{-6}$ radiative corrections to the effective potential of the
inflaton field may modify its shape at $\phi \sim M_{\rm P}$.
However, this problem does not appear if the flatness of the inflaton
potential is protected by supersymmetry.

Thus we assume the first step in reheating is to convert a fraction
$\delta$ of the inflaton energy density into a background of
baryon-number violating $X$ bosons.  They can be produced even if the
reheating temperature to be established at the subsequent stages of
reheating is much smaller than $M_X$. Here we see a significant
departure from the old scenario.  In the old picture production of $X$
bosons was kinematically forbidden if $M_\phi< M_X$, while in the new
scenario it is possible as the result of coherent effects. The
particles are produced out-of-equilibrium, thus satisfying one of the
basic requirements to produce the baryon asymmetry \cite{sak}.

The parametric resonance is efficient only if the $X$ lifetime is
greater than the typical time during which the number of $X$ bosons
grows $e$ times. During the stage of broad parametric resonance this
condition typically implies that the lifetime of the $X$ is greater
than about $10 M_\phi^{-1}$. Assuming the width for $X$ decay is
$\Gamma_X=\alpha_XM_X$, this requires $\alpha_X{\
\lower-1.2pt\vbox{\hbox {\rlap{$<$}\lower5pt\vbox {\hbox{$\sim$}}}}\
}10^{-2}$.  This is certainly true if  $X$-decay into
top quarks is kinematically forbidden. In the beginning of
reheating this condition is satisfied, e.g., if fermions acquire mass
greater than $M_X/2$ due to interaction with the inflaton field. At the
end of reheating the top quark mass receives a large  non-thermal
correction by means of the interaction with the $X$ bosons  \cite{inprep},
$m_t\sim h_t \langle X^2\rangle^{1/2} \sim h_t \lambda_\phi^{-1/4} 10^{15} $ GeV which is typically much greater than $M_X$.  Also, one can always
envisage the situation in which the $X$ boson generating the baryon
asymmetry does not belong to the same representation of the GUT group
which gives mass to the third generation.  Therefore, from now on we
will assume that the $X$ bosons may decay only to light fermions and
that they decay well after the end of explosive particle production,
resulting in a reheating temperature much smaller than $M_X$.

A self-interaction term in the Lagrangian of the type $\lambda_X
|X|^4$ also provides a non-thermal mass to the $X$ boson of the order
of $(\lambda_X \langle X^2\rangle)^{1/2}$, which we assume to be
smaller than the bare mass $M_X$, i.e., $\lambda_X{\
\lower-1.2pt\vbox{\hbox {\rlap{$<$}\lower5pt\vbox{\hbox{$\sim$}}}}\ }
10^{-2}\lambda_\phi^{1/2}$. However, this condition may be somewhat
relaxed since the parametric resonance may occur even if the effective
mass $M_X$ grows in its process, because the same happens to the
effective mass of the inflaton \cite{explosive}. Self-interactions do
not terminate the resonance effect since most particles remain inside
the resonance shell; furthermore creation of quanta different from
$X$, e.g., gauge bosons, are suppressed by kinematical reasons if the
non-thermal plasma mass of the final states is larger than the initial
energy of the $X$ particles. This happens if $\lambda^{1/2}_\phi{\
\lower-1.2pt\vbox{\hbox{\rlap{$<$}\lower5pt\vbox{\hbox{$\sim$}}}}\ }g$
\cite{inprep}, where we denote by $g$ the generic coupling constant
between the final states and $X$.

The next step in reheating is the decay of the $X$ bosons.  We assume
that the $X$ decay products rapidly thermalize.  It is only after this
point that it is possible to speak of the temperature of the universe.

The remaining energy in the inflaton is extracted in the final stage
of the reheating process.  After the parametric resonance period ends
and $X$ particle production shuts off, the inflaton performs small
oscillations around the minimum of the effective potential and the
universe soon becomes matter dominated.  A slow process of particle
production continues until the Hubble time becomes comparable to the
inflaton decay time, and the inflaton decays.  This part of the
picture is similar to the old reheating scenario.  Note that the above
estimate of $T_F$ did not depend upon the initial energy stored in the
inflaton, it only assumed that the energy of coherent oscillations
dominated the energy density.

As outlined above, we will consider a three part reheating process,
with initial conditions corresponding to the frozen universe at the
end of inflation.  The first stage is explosive particle production,
where a fraction $\delta$ of the energy density at the end of
preheating is transferred to $X$ bosons, with $(1-\delta$) of the
initial energy remaining in $\phi$ coherent oscillation energy. We
assume that this stage occurs within a few Hubble times of the end of
inflation.  The second stage is the $X$ decay and subsequent
thermalization of the decay products.  We assume that decay of an
$X$--${{\overline{X}}}$ pair produces a net baryon number $\epsilon$,
as well as entropy.  Reheating is brought to a close in the third
phase when the remaining energy density in $\phi$ oscillations is
transferred to radiation.

The description simplifies if we assume zero initial kinetic energy of
the $X$s.  This is a good approximation, since for small
$\lambda_\phi$ particles are produced with nonrelativistic
velocities. We also assume that there are fast interactions that
thermalize the massless decay products of the $X$.  Then in a
co-moving volume $a^3$, the total number of $X$ bosons, $N_X=n_Xa^3$,
the total baryon number, $N_B=n_Ba^3$, and the dimensionless radiation
energy, $R=\rho_Ra^4$, evolve according to
\begin{eqnarray}
\label{network}
\dot{N}_X & = & - \Gamma_X \left( N_X - N_X^{EQ} \right); \quad
 \dot{R}    =   - a M_X \dot{N}_X;
\nonumber \\
\dot{N}_B & = & - \epsilon \dot{N}_X
       - \Gamma_X N_B \left( N_X^{EQ} / N_0 \right) .
\end{eqnarray}
$N_X^{EQ}$ is the total number of $X$s in thermal equilibrium at
temperature $T \propto R^{1/4}$, and $N_0$ is the equilibrium number
of a massless degree of freedom in a comoving volume.

Fig.\ 1 shows the results of an integration of Eqs. (\ref{network}) in
a toy model with $M_\phi = 10^{13}$GeV, $M_X = 10^{14}$GeV, $\Gamma_X
= 5 \times 10^{-6} M_X$, $\Gamma_\phi = 5 \times 10^{-10} M_\phi$ ,
and two degrees of freedom ($b$ and ${{\overline{b}}}$).  Initial
conditions were chosen at $a=a_I$ to be $\rho_X = \rho_{\phi }
\sim 10^{-4} M_\phi^2 M^2_{\rm P}$, and $R = N_B = 0$.  The $\rho_X =
\rho_{\phi }$ assumption corresponds to $\delta=1/2$. Since the
number of $X$ bosons produced is proportional to $\delta$, the final
asymmetry is proportional to $\delta$.  A more quantitative
understanding of particle production in preheating is clearly
required.  However, we note that $B/\epsilon \sim 10^{-9}$ can be
obtained for $\delta$ as small as $10^{-6}$.

%%%%%%%%%%%%%%%%%%%%%%%%%%%%%%%%%%%
%%%%%%%%%%%%%%%%%%%%%%%%%%%%%%%%%%%
\begin{figure}
\centerline{ \epsfxsize=250pt  \epsfbox{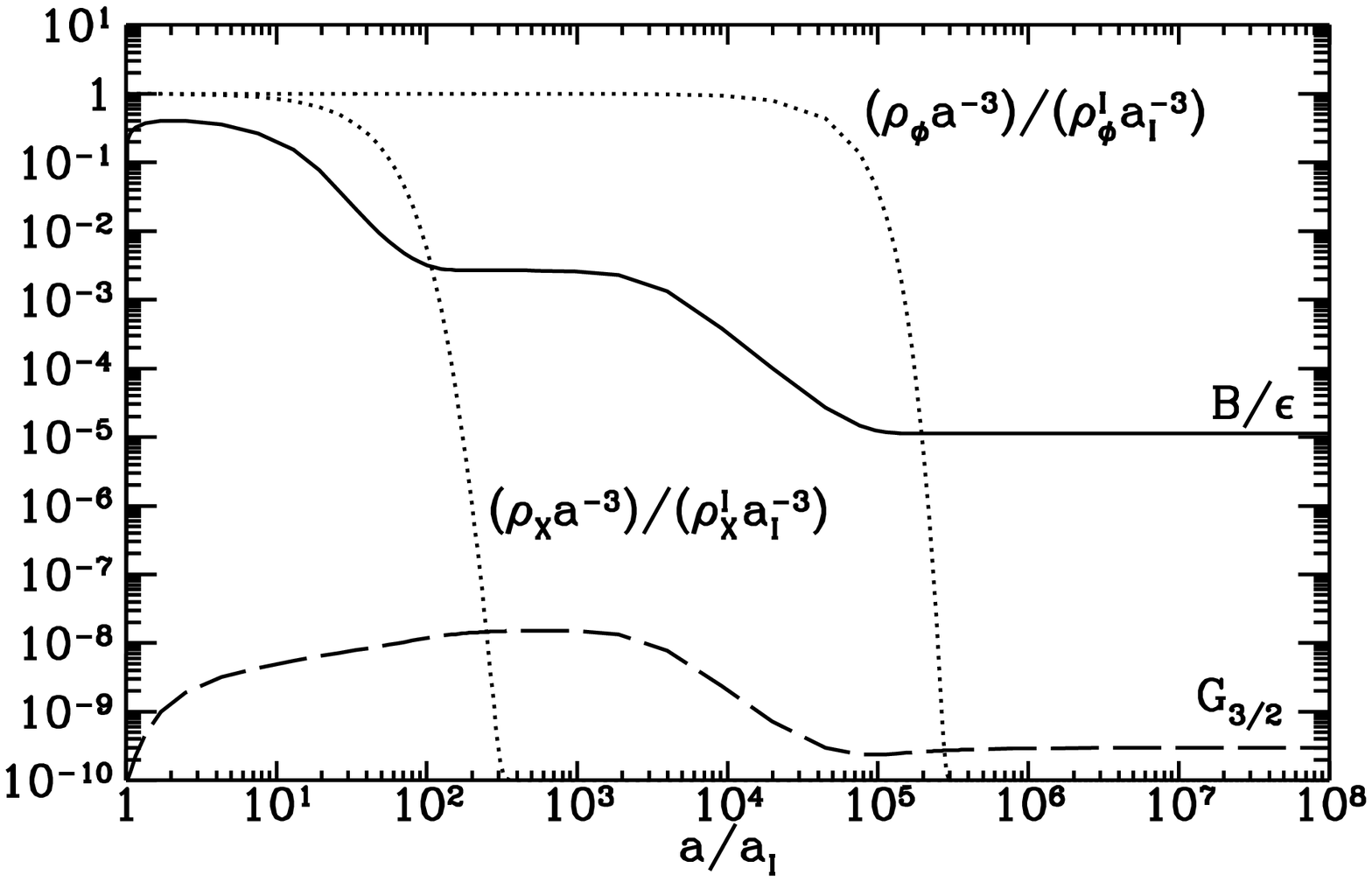} }
%\vspace*{3.5in}
\footnotesize{ \hspace{2em} Fig 1.  The evolution of the baryon
 number, the $X$ number density, the energy density in $\phi$
 oscillations, and the gravitino-to-entropy ratio as a function
 of the scale factor $a$.}
\end{figure}
%%%%%%%%%%%%%%%%%%%
%%%%%%%%%%%%%%%%%%%

The details of our scenario can be altered by many factors. For
example, when the density of $X$ particles decreases in expanding
universe, the effective mass of the top quark also decreases, which
may open the possibility of $X$ decay to top quarks. Therefore at some
moment the decay rate of the $X$ bosons may suddenly increase. This
will change some of our numerical results \cite{inprep}. However, we
believe that our simple model demonstrates the general behavior that
might be expected in more realistic/complicated models.  The baryon
number $B = n_B/s$ rapidly rises.  However $B$ decreases as entropy is
created and $X$ inverse reactions damp the baryon asymmetry.  After
most of the energy is extracted from the initial $X$ background, the
baryon number is further damped as entropy is created during the decay
of energy in the $\phi$ background.  In the model illustrated in Fig.\
1, the final value of $B/\epsilon$ is $5\times10^{-4}$.

We have numerically integrated the equation governing the number
density of gravitinos $n_{3/2}$ \cite{limit}. The result for
$G_{3/2}=n_{3/2}/s$ is shown in Fig.\ 1. Notice that, even though
gravitinos are copiously produced at early stages by scatterings of
the decay products of the $X$, $G_{3/2}$ decreases as entropy is
created during the subsequent decay of energy in the $\phi$
background.  A similar behavior has been found in
\cite{japan}. Successful nucleosynthesis requires $G_{3/2}{\
\lower-1.2pt\vbox{\hbox {\rlap{$<$}\lower5pt\vbox{\hbox {$\sim$}}}}\
}10^{-10}$ which translates into an upper bound on the inflaton decay
rate, $\alpha_\phi{\ \lower-1.2pt\vbox{\hbox
{\rlap{$<$}\lower5pt\vbox{\hbox {$\sim$}}}}\ } 10^{-10}$.

As $X$ particles decay long after the end of the stage of preheating
and their energy density is considerably diminished by the expansion
of the Universe, the maximum of the thermalization temperature of
their decay products is considerably smaller than the unification
scale $10^{16}$ GeV.  This means that GUT symmetry is not restored
when $X$ decay products thermalize. If not the case, the subsequent
decrease of the temperature of the thermal bath would be accompanied
by a GUT symmetry breaking phase transition and the generation of
dangerous topological defects. For the same reason, we require that
GUT symmetry is not restored at the early stages of preheating, when
non-thermal effects are dominant \cite{KLSSR,tkachev,tr}. Let $\Phi$
be the field responsible for GUT symmetry breaking with a potential of
the form $V(\Phi)=-\mu^2\Phi^2+\lambda_\Phi \Phi^4$, $\mu\sim 10^{16}$
GeV.  The $X$ boson may couple to the $\Phi$ by an interaction of the
type $\lambda |\Phi|^2|X|^2$. This interaction induces a mass squared
for the $\Phi$ field of order of $\lambda\langle X^2\rangle\sim
10^{-2}\lambda\lambda_\phi^{-1/2} M_{\rm P} M_\phi$ \cite{KLSSR}. This
term is smaller than $\mu^2$ and does not lead to symmetry restoration
for $\lambda{\ \lower-1.2pt\vbox{\hbox {\rlap{$<$}\lower5pt\vbox{\hbox
{$\sim$}}}}\ }10^{2}\lambda_{\phi}^{1/2}$ \cite{inprep}. This
condition is not difficult to satisfy. Therefore parametric resonance
does not lead to GUT phase transitions and to the primordial monopole
problem in our scenario.

In conclusion, we have shown that the present baryon asymmetry may be
produced after inflation in the decay of non-thermal GUT bosons
produced in preheating. Our scenario solves many of the serious
shortcomings of GUT baryogenesis in the old theory of reheating where
it was kinematically impossible to produce superheavy particles after
inflation. The out-of-equilibrium condition is naturally attained when
superheavy quanta are produced in the regime of broad parametric
resonance after the stage of inflation and considerably differs from
the out-of-equilibrium condition in the GUT thermal scenario
\cite{review} where superheavy bosons decouple from the thermal bath
when relativistic if $K=(\Gamma_X/H)_{T=M_X}\ll 1 $ and then decay
producing the baryon asymmetry.  Gravitinos are subsequently diluted
by the entropy released during the late decay of the inflaton field
and their abundance can be easily accommodated to be in agreement with
the successful predictions of nucleosynthesis.

Our scenario is based on several assumptions about the structure of
the theory and on relations between various coupling constants.  For
the parameters used to generate the results of Fig.\ 1, baryon number
generation was relatively efficient: $B/\epsilon\sim 5\times10^{-4}$.
Within uncertainties of model parameters, the value of $\epsilon$,
etc., the present $B\sim10^{-10}$ may arise from GUT baryogenesis
after preheating.  Of course, additional work is needed to implement
the ideas discussed above in the context of a more realistic model.
However, we feel very encouraged that  recent progress in the
theory of reheating has removed many obstacles which precluded
successful GUT baryogenesis in inflationary cosmology.  We will
present more details in a subsequent publication \cite{inprep}.

E.W.K. and A.R. acknowledges the organizers of the Astroparticle
workshop held in Uppsala, Sweden, where part of this work was done. We
would like to thank Chris Hill, Lev Kofman and Alexei Starobinsky for
fruitful discussions.  A.L.\ is supported in part by NSF Grant No.\
PHY--8612280.  E.W.K.\ and A.R.\ are supported by the DOE and NASA
under Grant NAG5--2788.

\end{document}